\documentclass[sigconf]{acmart}

\AtBeginDocument{%
  }

\makeatletter
\def\@ACM@checkcolwidth{}
\makeatother

\copyrightyear{2025}
\acmYear{2025}
\setcopyright{rightsretained}
\acmConference[CHI EA '25]{Extended Abstracts of the CHI Conference on
Human Factors in Computing Systems}{April 26-May 1, 2025}{Yokohama, Japan}
\acmBooktitle{Extended Abstracts of the CHI Conference on Human Factors in
Computing Systems (CHI EA '25), April 26-May 1, 2025, Yokohama,
Japan}\acmDOI{10.1145/3706599.3720266}
\acmISBN{979-8-4007-1395-8/2025/04}

\acmConference[CHI EA '25]{Make sure to enter the correct
  conference title from your rights confirmation emai}{April 26--May 01,
  2025}{Yokohama, Japan}
\acmISBN{978-1-4503-XXXX-X/18/06}

\usepackage{float}
\usepackage{framed}
\usepackage{enumitem}
\usepackage{calc}
\usepackage{booktabs}
\usepackage{multirow}
\usepackage{geometry} 
\usepackage{subcaption}
\usepackage{listings}
\usepackage{xspace}

\newcommand{\Sys}{\textsc{PolicyPulse}\xspace}

\title[\Sys]{\Sys: LLM-Synthesis Tool for Policy Researchers}
\subtitle{
Human-AI Tool for Synthesizing Public Experiences about Policy-Relevant Topics into Themes }

\author{Ella Colby}
\authornotemark[1]
\affiliation{%
  \institution{Princeton University}
  \city{Princeton}
  \state{New Jersey}
  \country{USA}
  \postcode{08540}
}
\email{ellacolby@princeton.edu}

\author{Jennifer Okwara}
\authornotemark[1]
\affiliation{%
  \institution{Princeton University}
  \city{Princeton}
  \state{New Jersey}
  \country{USA}
  \postcode{08540}
}
\email{jo4986@princeton.edu}

\author{Maggie Wang}
\authornote{All authors contributed equally to this research.}
\affiliation{%
  \institution{Princeton University}
  \city{Princeton}
  \state{New Jersey}
  \country{USA}
  \postcode{08540}
}
\email{maggiewang@princeton.edu}

\author{Varun Nagaraj Rao}
\affiliation{%
  \institution{Princeton University}
  \city{Princeton}
  \state{New Jersey}
  \country{USA}
  \postcode{08540}
}
\email{varunrao@princeton.edu}

\author{Yuhan Liu}
\affiliation{%
  \institution{Princeton University}
  \city{Princeton}
  \state{New Jersey}
  \country{USA}
  \postcode{08540}
}
\email{yuhanl@princeton.edu}

\author{Andr{\'e}s Monroy-Hern{\'a}ndez}
\affiliation{%
  \institution{Princeton University}
  \city{Princeton}
  \state{New Jersey}
  \country{USA}
  \postcode{08540}
}
\email{andresmh@princeton.edu}

\renewcommand{\shortauthors}{Colby, Okwara, and Wang et al.}

\begin{abstract}
Public opinion shapes policy, yet capturing it effectively to surface diverse perspectives remains challenging. This paper introduces \Sys, an LLM-powered interactive system that synthesizes public experiences from online community discussions to help policy researchers author memos and briefs, leveraging curated real-world anecdotes. Given a specific topic (e.g.,  ``Climate Change''), \Sys returns an organized list of themes (e.g.,  ``Biodiversity Loss'' or ``Carbon Pricing''), supporting each theme with relevant quotes from real-life anecdotes. We compared \Sys outputs to authoritative policy reports. Additionally, we asked 11 policy researchers across multiple institutions in the Northeastern U.S to compare using PolicyPulse with their expert approach. We found that \Sys's themes aligned with authoritative reports and helped spark research by analyzing existing data, gathering diverse experiences, revealing unexpected themes, and informing survey or interview design. Participants also highlighted limitations including insufficient demographic context and data verification challenges. Our work demonstrates how AI-powered tools can help influence policy-relevant research and shape policy outcomes.
\end{abstract}



\begin{document}
\begin{teaserfigure}
  \centering
  \includegraphics[width=0.7\textwidth]{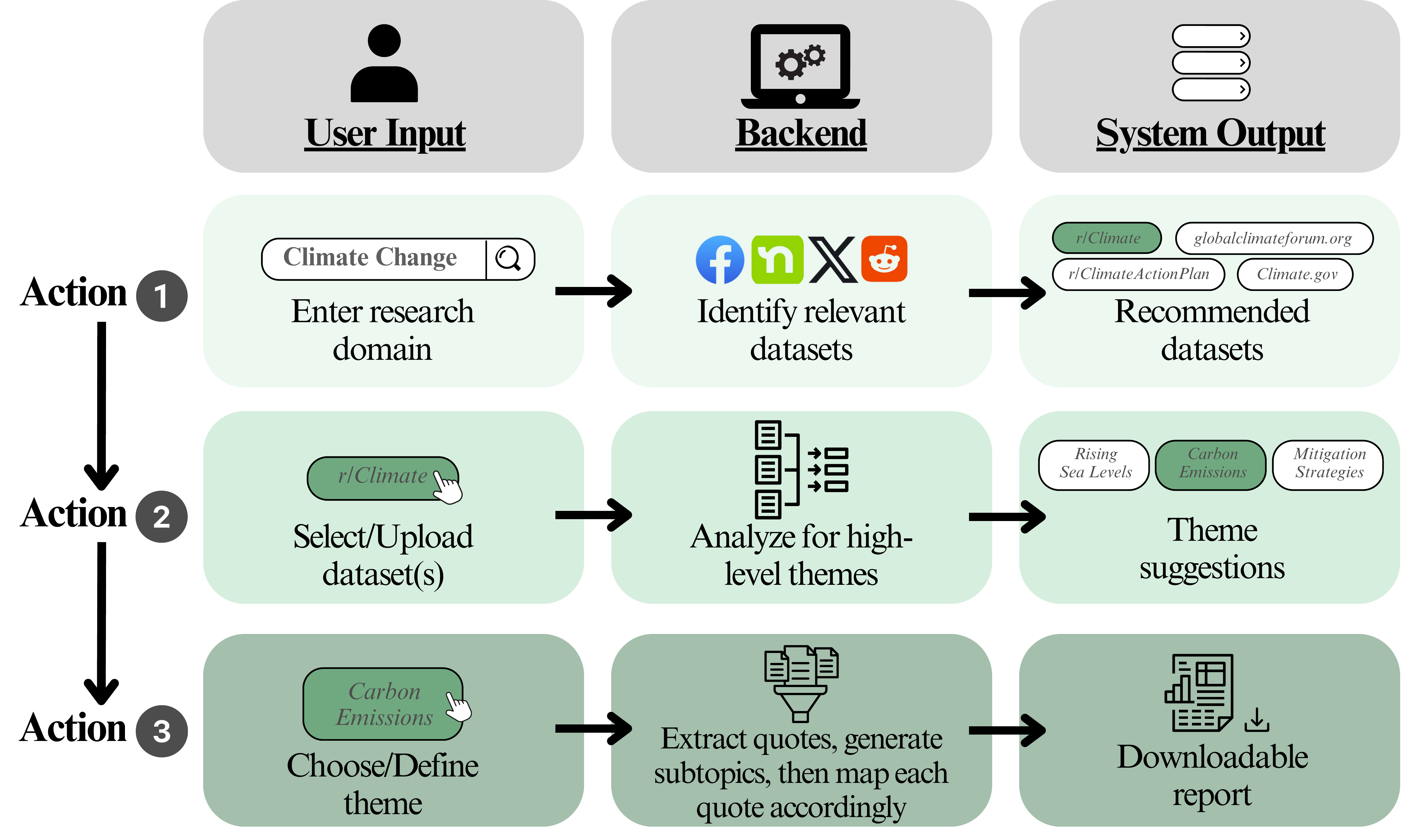} 
  \caption{We illustrate \Sys's three‐step workflow. Action 1: The user enters a policy-relevant topic (left), prompting the backend (center) to identify relevant datasets, which the system (right) then presents as recommendations. Action 2: The user selects or uploads the dataset(s), triggering the backend to analyze for high‐level themes, and the system outputs theme suggestions. Action 3: The user chooses or defines a final theme, prompting the backend to extract relevant quotes, generate subtopics, and map each quote accordingly, resulting in a downloadable report.}
  \label{fig:SystemArch}
\end{teaserfigure}

\maketitle


\section{Introduction}
Capturing real-world public experiences is essential for shaping policies that reflect diverse societal perspectives \cite{stimson2018public}. Policy researchers rely on three main data sources to create policy memos and briefs \cite{engler2020all}: primary sources like surveys and interviews, secondary sources such as government databases (e.g., Bureau of Labor Statistics, National Institutes of Health in the U.S.) and think tank reports (e.g., Pew Research), and microsimulations that devise computational models to measure policy effects. However, these traditional approaches often fail to capture diverse public perspectives, particularly from underserved communities who lack time or resources to participate in formal consultations \cite{landemore2021opendemocracy, jacobs2011oxford, schulman2024methods}. The U.S. Office of Management and Budget's (OMB) recent initiative on Public Participation and Community Engagement (PPCE) \cite{schulman2024methods} highlights this critical need to incorporate more diverse voices in policy discussions. Online communities on platforms like Reddit offer rich, candid discussions on policy-relevant topics \cite{rao2024rideshare, fiesler2024remember} but are underutilized to impact policy due to the challenges of synthesizing unstructured data.

To help mitigate these challenges, we present \Sys, an LLM-powered interactive system that synthesizes online community discussions to assist policy researchers -- obtaining a ``pulse'' on public experiences. We explore three key research questions: (1) \textit{How can LLM-driven tools be designed to surface people's real-life experiences and anecdotes on policy-relevant topics?} (2) \textit{What approaches can effectively enable policy researchers to leverage these tools?} and (3) \textit{To what extent do LLM-driven tools benefit policy researchers compared to typical data sources?} To answer these questions, we: 
\begin{itemize}
    \item devised a multiphase prompting strategy to extract relevant experiences from online discussions inspired by recent advances in LLMs \cite{rao2024quallm, pham2024topicgpt, lam2024concept}
    \item engineered an interactive interface for policy researchers to explore themes derived from these anecdotes
    \item evaluated how \Sys enhances secondary data analysis while complementing primary data collection by informing survey design
\end{itemize}

For our mixed methods evaluation of \Sys, we chose two policy-relevant topics: \textit{``Climate Change''} and \textit{``Social Media and Kids''} -- to assess its ability to complement primary methods (e.g.,  surveys) and secondary sources (e.g.,  authoritative reports). \Sys captured 73\% and 84\% of the themes identified in the authoritative reports \cite{wmo2021, pew2020}, respectively, while quickly uncovering unique insights such as emerging trends and anecdotes. Participants in our user study (N=11) praised the diversity of themes uncovered, the tool’s ability to reveal unexpected perspectives and kickstart their research, particularly when exploring unfamiliar domains, and the reduced costs compared to surveys and listening sessions. They also noted areas for improvement, including enhanced data verification and demographic context. Ultimately, our work demonstrates how AI tools can democratize access to public opinions by surfacing diverse voices that might not participate in traditional forums and by responding to calls to inform workflows for integrating system-people-policy interactions \cite{yang2024future, jackson2014policy, lazar2015public, davis2012occupy, spaa2019understanding}.

\begin{figure*}[htb]
  \centering
  \includegraphics[width=0.8\textwidth]{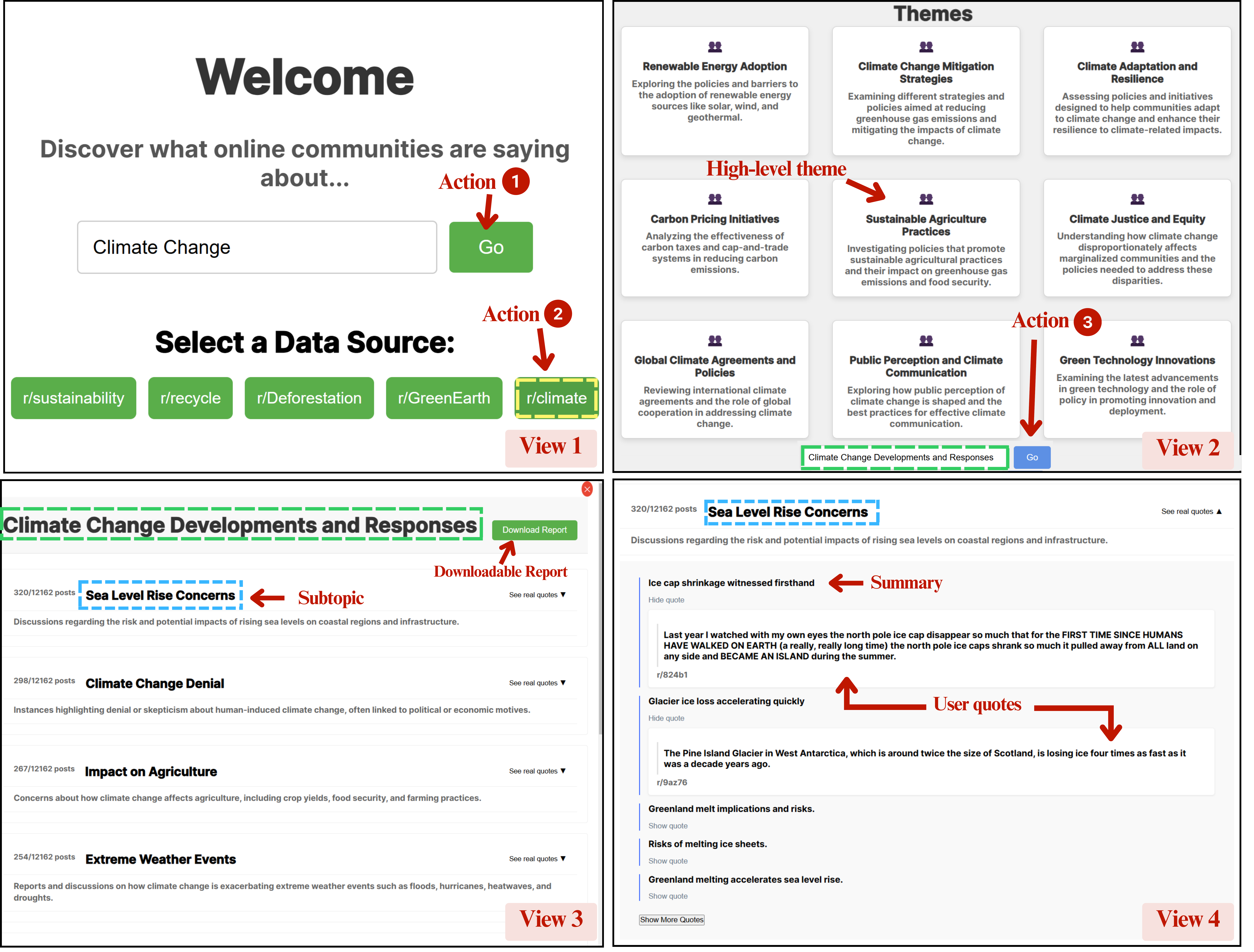}  
  \caption{View 1 shows the interface. This view prompts users to input a research domain for analysis (Action 1) and select a data source (Action 2). View 2 shows high-level themes in the \Sys interface. The user can then select or search for a primary research topic (Action 3). View 3 shows the final Report View of the interface, displaying all subtopics identified from the data source and quote counts. View 4 shows examples of 5-6 word summaries beneath the subtopic and a couple of fully displayed quotes on which the summaries are based.}
  \label{fig:Screens}
\end{figure*} 

\section{BACKGROUND AND RELATED WORK}
\subsection{The Policy Research Process}
Policy researchers create policy memos \cite{harvard2024memo} and briefs \cite{unc2024memo} to influence public policy by providing concise, evidence-based analyses of policy issues and actionable recommendations. Producing these documents requires effectively gathering and utilizing data from three main sources \cite{engler2020all}: (1) Primary sources, such as surveys, interviews, listening sessions, sensors, and other tools for firsthand data collection; (2) Secondary sources, including analyses of existing data like government databases (e.g.,  U.S. National Institutes of Health and Bureau of Labor Statistics), academic publications, and think tank reports (e.g.,  Pew Research); and (3) Microsimulations, computational models that simulate the effects of policy changes on individuals, households, or firms. PolicyPulse enhances this process by analyzing secondary data (e.g.,  online community forums) and complementing primary data collection by informing survey or interview design. Responding to calls from HCI and CSCW scholars \cite{yang2024future, jackson2014policy, lazar2015public, davis2012occupy, spaa2019understanding}, PolicyPulse also demonstrates the integration of system-people-policy interactions to improve policy outcomes.

\subsubsection{Bridging Public Participation and Policy}
The U.S. federal government launched a Request for Information (RFI) in 2024 to develop government-wide frameworks for Public Participation and Community Engagement (PPCE), focusing on underserved communities \cite{schulman2024methods}. This initiative highlights the need for tools that efficiently gather and analyze public opinion for policy research. Social media and public discussion forums offer valuable platforms to enhance PPCE by incorporating diverse perspectives, including traditionally underrepresented voices. While synthesizing these varied viewpoints remains challenging, \Sys demonstrates one approach through its LLM-powered analysis of online discussions.

\subsection{LLMs for Thematic Analysis} 
Recent advances in LLMs have enabled new approaches to analyzing unstructured text, such as online discussions on forums like Reddit. Tools like TopicGPT, DocETL, and LLooM \cite{pham2024topicgpt, lam2024concept, shankar2024docetlagenticqueryrewriting} can extract topics from large text corpora, but they typically require technical expertise in prompt engineering, API integration, and programming. QuaLLM \cite{rao2024quallm} demonstrated the potential for LLMs to analyze online discussion forums and extract themes through a multiphase prompting approach, yet lacks an accessible system for users. Other LLM-wrapper tools such as Reddit Answers \footnote{\url{https://www.reddit.com/answers/}} and The Giga Brain\footnote{\url{https://thegigabrain.com/feed}} have made search more accessible, but don't allow users control over the actual data sources (e.g., which subreddits).

We build on these works by rethinking how LLM-powered analysis tools can be accessible to policy researchers. Rather than requiring users to write code or craft prompts, PolicyPulse encapsulates the complexity of LLM operations behind a visual, intuitive interface, while still allowing users significant agency over data sources to gather real public anecdotes and experiences on policy-relevant topics.

\section{POLICYPULSE SYSTEM DESIGN}

\begin{figure*}[htb]
  \centering
  \includegraphics[width=\textwidth,trim={0 0 0 0},clip]{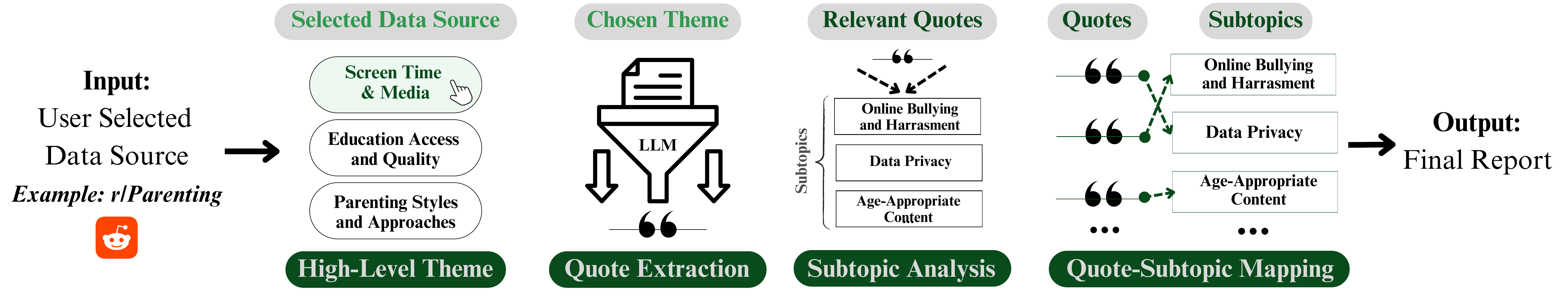}
  \caption{These four prompts drive the system from raw data source to final reporting. First, the user selects or provides a raw text data source. The High-Level Theme Prompt proposes high-level research themes or accepts custom ones. Given the chosen theme, the Quote Extraction Prompt collects relevant quotes and highlights anecdotes to reduce bias. Next, the Subtopic Analysis Prompt identifies subtopics within these quotes. Finally, the Mapping Prompt assigns each quote to its most relevant subtopic, generating a structure for the final report.}
  \label{fig:prompts}
\end{figure*}

\Sys transforms online discussions into structured insights for policy research through a three-stage analysis pipeline.

\subsection{Interactive Workflow}
Users interact with PolicyPulse through three main stages:
\subsubsection{Data Source Selection}
Users specify a policy domain (e.g.,  ``Climate Change'') and select relevant data sources. The system uses an LLM to match topics with policy-relevant online communities, currently providing options from subreddit forums but designed for expansion to other platforms and alternative data sources (Figure~\ref{fig:Screens}, View 1).
\subsubsection{Theme Generation}
The system identifies high-level themes within the selected data source using LLM-powered analysis. Users can explore suggested themes or perform custom searches based on their own research needs (Figure~\ref{fig:Screens}, View 2).
\subsubsection{Report Generation}
For each selected or entered theme, PolicyPulse processes data through a series of specialized LLM prompts (Figure~\ref{fig:prompts}). This multi-stage pipeline ensures structured, concise insights tailored for policy research. Specifically:
\begin{itemize}
    \item Relevant quotes of people's actual experiences and anecdotes are extracted using a Quote Extraction prompt designed to minimize bias (Figure~\ref{fig:prompts}: Prompt 2).
    \item Aggregated quotes are analyzed to identify subtopics (Figure~\ref{fig:Screens}: View 3, Figure~\ref{fig:prompts}: Prompt 2).
    \item Quotes are mapped to appropriate subtopics to ensure structured organization (Figure~\ref{fig:prompts}: Prompt 3).
    \item Concise summaries (5–6 words) are generated for enhanced readability (Figure~\ref{fig:Screens}: View 4).
    \item A final downloadable report is created for offline analysis.
\end{itemize}

Our prompt engineering approach ensures consistent, high-quality outputs while maintaining a policy research focus (detailed prompts in Appendix~\ref{appendix:prompting}). The system processes approximately 1,000 quotes per 10 minutes, caching results for immediate access on subsequent views.
\subsection{Technical Implementation}
The system uses a RESTful API to orchestrate interactions between the frontend and LLM-based analysis pipeline. Core functionality includes data preprocessing through Pandas, efficient caching for iterative analysis, and JSONL-based storage for structured reports. While currently leveraging The Eye\footnote{\url{https://the-eye.eu/redarcs/}} Reddit archive, the modular architecture supports structured processing of any user-uploaded dataset. Furthermore, this allows the user agency over data sources, which current LLM-wrapper tools such as Reddit Answers\footnote{\url{https://www.reddit.com/answers/}} and The Giga Brain\footnote{\url{https://thegigabrain.com/feed}} do not allow. This is especially important for policy researchers looking to explore public sentiment using only specific data sources they deem to satisfy their requirements. 

\subsubsection{Data Collection, Prompting, and Storage}
\Sys currently relies on The Eye Reddit archive as its primary data source. Due to its vast user base, Reddit is a valuable source for gauging public sentiment on policy-relevant topics. Its anonymity-driven candid discussions and specialized subreddit communities facilitate rich contextual data and longitudinal analysis of public opinion \cite{fiesler2024remember, rao2024rideshare, chen2021using, xu2024public, huang2024politically}.  Furthermore, comprehensive data is easily and publicly accessible. The initial script that handles downloaded data scraped from The Eye aggregates posts and their associated comments into unified discussion threads to preserve contextual coherence, ensuring that analyses accurately reflect the temporal order of public discourse. The output raw data is stored in CSV format to optimize access and preprocessing efficiency. The Flask backend processes this raw data into JSON files categorized by prompting stages (Figure~\ref{fig:prompts}). This organized process ensures traceability and facilitates structured data storage and retrieval, drawing upon the four-stage multi-prompting strategy from QuaLLM with modifications for streamlined generalization. Emphasizing a stepwise policy research workflow, each prompt is carefully engineered to maintain consistency, ensure high-quality outputs, mitigate bias, and support policy-focused analytical objectives. As shown in Appendix~\ref{appendix:prompting}, we detail our prompt engineering approach for each system component.

\section{METHODS}

\begin{figure*}[htb]
    \centering
    \includegraphics[width=0.8\linewidth]{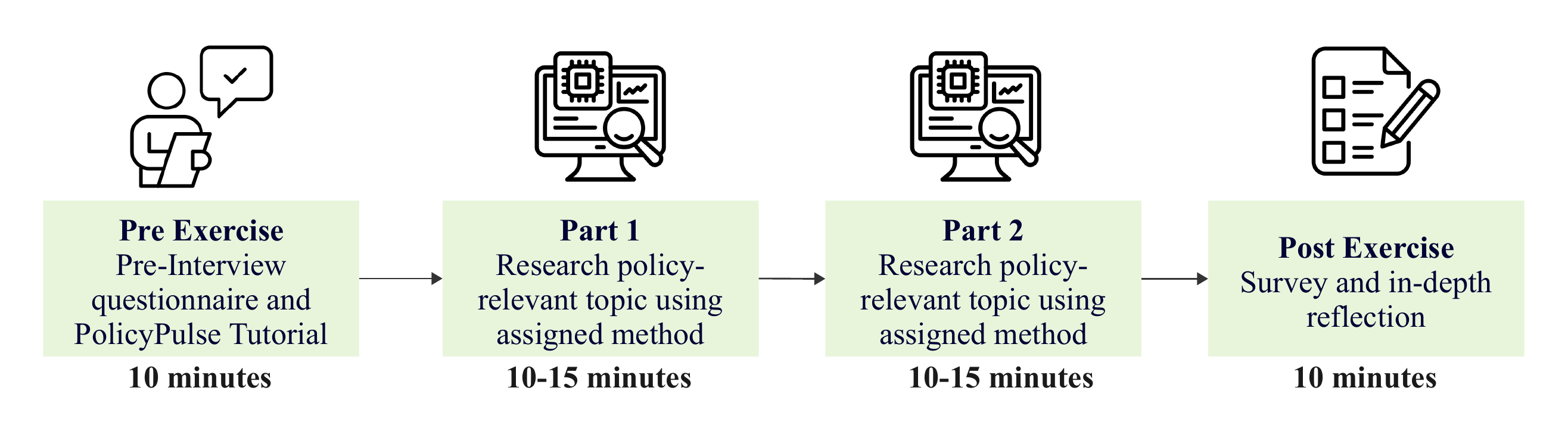}
    \caption{Timeline of the User Study. This figure illustrates the four-phase research method comparing AI-assisted and participants' own non-AI expert approach. The process begins with an interview and tutorial, followed by two sequential research parts randomized based on topic (Climate Change or Social Media \& Kids) and method (with \Sys or own non-AI expert approach), and concludes with a retrospective survey and interview. Each research phase is allocated 10-15 minutes, with participants documenting both quantitative and qualitative findings using standardized interviewee report templates throughout the process.}
    \label{fig:timeline}
\end{figure*}

We evaluated \Sys through a mixed-methods analysis consisting of quantitative data analysis, interviews, and surveys. 
Initially, we conducted a comparative analysis of \Sys output reports alongside authoritative reports (see \ref{section:Auth_Report}). We selected two policy-relevant issues: ``Climate Change'' and ``Social Media and Kids.'' \cite{wmo2021, pew2020}. Two researchers independently performed thematic analysis on both sets of authoritative reports, using a combination of their own interpretation and explicitly listed themes and topics. They reached consensus on a final list of themes for each report. The researchers then mapped themes from the authoritative reports to those generated by \Sys, calculating the percentage of coverage. Detailed themes and mappings are presented in Appendix \ref{appendix:auth_comparison}.

Having established the initial feasibility and validity of the system, we then recruited 11 experienced policy experts, including policymakers, researchers, and professors, to evaluate \Sys through comparative analysis with participants' own non-AI expert approaches. During 45-60-minute interviews, participants first completed a pre-task survey assessing their policy research experience, followed by a structured comparison task. Participants were randomly divided into two groups based on the order of topics they would research. Within each group, they were further divided based on the order of methods used (\Sys vs. participants' own non-AI expert approach). This organizational structure helped to control for order effects, such as practice or fatigue, which could influence participants' performance depending on the sequence of tasks. Each participant spent 10-15 minutes researching one of two randomly assigned topics using one method and then switched to the other method for the second topic. For both topics, they recorded themes and anecdotal evidence in standardized worksheets (see Appendix ~\ref{sec:Interview Worksheet}), enabling a direct comparison of research efficiency and effectiveness.

We audio recorded and transcribed all interviews, then conducted thematic analysis using open coding. We coded the transcripts to identify recurring themes and patterns in user feedback. We iteratively refined these codes through discussion until reaching consensus, then grouped them into higher-level themes. This analysis revealed three primary themes: enhancement of traditional methods, interface design benefits, and areas for improvement.

Post-task Likert-scale surveys measured user experience and perceived benefits across multiple dimensions, including speed, breadth of perspectives, anecdote discovery, data quality, and ease of analysis. Participants also reflected on how \Sys compares to and may complement surveys, interviews, and listening sessions. We quantitatively analyzed the worksheet data to compare the number of themes gathered within the time constraint across both methods. The study was approved by our institution's IRB. 

\section{RESULTS}

\begin{figure*}[htb]
  \centering
  \includegraphics[width=0.6\textwidth]{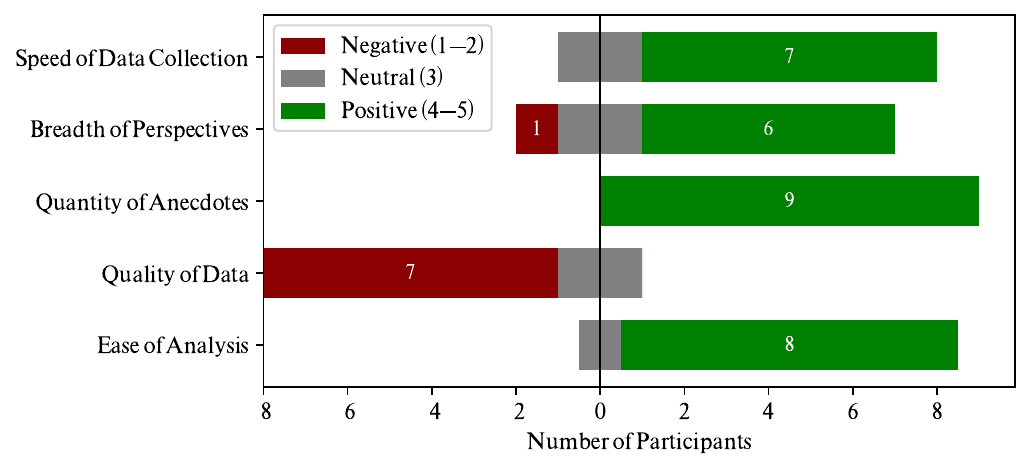}  
  \caption{This diverging stacked bar chart breaks down Likert scale responses (1-5) into three groups (negative, neutral, positive) with the y-axis representing one question from our survey. The length of each colored segments indicates how many of our evaluation participants chose that group. The further left the red bar extends, the more negative; the further right the green bar extends the more positive. The gray area indicates the number of neutral participants. Notably, most participants found the quantity of anecdotes and ease of analysis provided by \Sys a key strength of the tool, but the data quality a major place for improvement.}
  \label{fig:sentimentchart}
\end{figure*}

\subsection{\Sys Outputs Align with Authoritative Reports While Offering Additional Unique Insights}
\label{section:Auth_Report}

\Sys effectively captures and complements key themes identified in authoritative reports while providing unique insights. Our comparative analysis revealed that for the ``Climate Change'' topic, \Sys covered 73\% (11/15) of the themes presented in the WMO report\cite{wmo2021}, and for ``Social Media and Kids'', it covered 84\% (16/19) of the themes from the Pew Research report\cite{pew2020} (See Appendix~\ref{appendix:auth_comparison} for detailed themes). This coverage demonstrates \Sys's ability to accurately represent key policy concerns while supplementing them with real-world anecdotes. More importantly, these findings serve as a validation of \Sys's functionality rather than emphasizing its accuracy in reproducing expert-identified themes. The system performed better for the more focused ``Social Media and Kids'' topic compared to the broader ``Climate Change'' topic, which demonstrates that \Sys performs better when given a more specific topic. This is likely due to how specific and granular user quotes are.

\Sys captures a wider and more diverse perspective often overlooked by traditional sources, offering real-time public sentiment and emerging trends. While an authoritative report for the ``Social Media and Kids'' topic may claim ``Vast majority of parents say parents and guardians have a lot of responsibility in protecting children from inappropriate content online''\cite{pew2020}, \Sys provides explicit examples from real parents. For instance, one Reddit user's quote in a \Sys report states: \textit{``I've been thinking a lot about raising my kid (not yet born) in the digital world and decided that I will not buy them a modern computer or a game console until he/she is a teen''}. This raises concerns about digital literacy, education, and child development, highlighting the need for balanced policies that provide technology access while mitigating potential risks to children's learning and development.

\Sys also captures nuances that may be absent in authoritative reports. For example, a teen's reluctance to report online bullying is explained by one Reddit user: \textit{``There's little to no chance that an early teen will tell you about bullying no matter how great your relationship is. They're horrendously embarrassed by it and it's also usually unexpected...you can have the most popular kid in her grade, and the next day everyone will turn on her.''} Such insights could potentially inform policy formation around encouraging anonymous reporting and expanding access to mental health resources.

The ability to easily re-analyze data enables researchers to capture recent developments and evolving public opinion efficiently. This feature, combined with \Sys's comprehensive coverage and unique insights, positions it as a valuable complement to traditional sources in policy research. Future evaluations could scale up to provide more comprehensive insights into \Sys's performance across various topic scopes, further validating its functionality and value in policy analysis.

\subsection{\Sys Enhances Traditional Research Methods While Reducing Resource Requirements}

\Sys complements existing policy research methods by providing a cost-effective and time-efficient solution. Figure \ref{fig:sentimentchart} demonstrates that 7/11 (64\%) of participants rated \Sys's speed and efficiency positively. Furthermore, participants on average collected 2 more themes during the study period compared to their own non-AI expert approach as seen in Figure \ref{fig:bargraph}. With regards to costs, participants told us that listening sessions and surveys cost between \$4,000-\$80,000 respectively, while \Sys can operate at much more reduced cost. Beyond financial benefits, P2 also told us the tool accelerates the public opinion gathering stage, which comprises 10\% of the policy research process: \textit{``Your tool is definitely faster than what we could produce in a survey because we would be surveying hundreds of people'' (P2)}.

\Sys also excels at capturing a broader range of public opinions compared to traditional methods. While conventional approaches exhibit inherent biases toward participants with time and resources, \Sys leverages online forums to expand demographic representation. One participant highlighted this advantage, stating \textit{``Most issues are not very researched in terms of polls and surveys. So this tool could become more versatile'' (P6)}. However, two users noted potential population bias favoring younger, internet-savvy generations, supporting \Sys's role as a supplementary rather than a replacement tool.

All participants believe \Sys integrates effectively into existing research workflows, particularly in early stages. Specifically, they explained the typical policy research pipeline, which starts with background research, consulting existing surveys from think tanks like Pew Research or other governmental agencies and databases, engaging subject matter experts, and finally conducting new surveys or listening sessions -- a process spanning 3-4 months. Participants felt that our tool proved especially valuable for researchers less familiar with topics, as its thematic layout facilitates rapid understanding. As one participant commented, \textit{``I would use your tool to jump start the research process'' (P8).}

\subsection{AI Integration and Interface Design Facilitate Objective Data Analysis}
Despite initial AI skepticism, participants found that \Sys's AI backend actually reduces bias in data presentation. Specifically, they felt the thematic categorization limits selection bias, particularly benefiting researchers already familiar with topics. And, the interface effectively presents qualitative data, making non-statistical information more accessible. One participant emphasized this benefit: \textit{``The surveys would say X percent of parents think that students shouldn't use social media. But then you wouldn't get more of the reasoning behind. And I've struggled with this in my previous position in policy research'' (P10)}.

The raw user data, presented as Reddit quotes, distinguishes \Sys from non-AI expert research. One participant noted, \textit{``So you know the sense I'm getting is that your tool [is] actually giving us...a little insight into what people are actually saying, as opposed to what people are summarizing about it'' (P2)}. The interface's efficiency impressed participants, with one stating, \textit{``I didn't have to read a bunch of stuff that I wasn't interested in reading'' (P11)}.

\begin{figure}[tb]
  \centering
  \includegraphics[width=0.6\columnwidth]{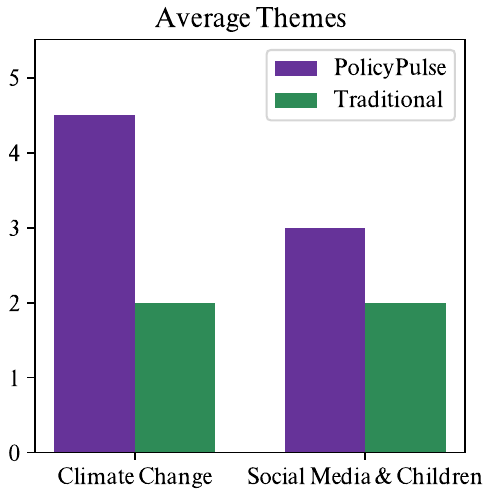}  
  \caption{This bar chart shows the average number of themes gathered across the two research topics---Climate Change and Social Media \& Kids---using two research methods, \Sys and non-AI expert approach. As we can see, even in a limited study duration, using \Sys, on average, allowed participants to collect a higher number of themes for both topics, pointing to the process being an order of magnitude faster.}
  \label{fig:bargraph}
\end{figure}

The tool also excels at revealing unexpected insights compared to participants' own non-AI expert methods. As one participant observed, \textit{``For unexpected themes I think it's much easier with your tool than with traditional methods'' (P7)}. They also felt that the AI backend facilitates access to otherwise hard-to-reach data and anecdotes, enhancing both accessibility and perspective diversity. The Likert scale responses in Figure \ref{fig:sentimentchart} validate these qualitative observations, with 8/11 (73\%) of participants rating \Sys's ease of analysis positively. Additionally, participants consistently gathered more themes using \Sys across both research topics (Figure \ref{fig:bargraph}), supporting their feedback about improved access to diverse data.

\subsection{Areas for Enhancement Focus on Metadata and Trust Building}
\label{area for improvement}
Participants identified several opportunities for improvement, primarily centered around metadata enhancement. These concerns are reflected in the quantitative data, where data quality received the lowest positive ratings among all measured dimensions in Figure \ref{fig:sentimentchart}. Diving deeper through retrospective interviews, participants desired demographic information including geographic, racial, and gender breakdowns to assess dataset representation. As one participant suggested, \textit{``Do try to categorize Reddit users into some of those demographic data...that's the kind of thing that I would be really interested in'' (P4)}.

The interface requires refinement in data presentation. Participants requested more intuitive quantitative representations, with one noting, \textit{``Sometimes it's helpful to have kind of the quantitative score of how that theme ranks, or what percentage'' (P7)}. Additional suggestions include improved scrolling functionality, reduced quote redundancy, and clearer theme hierarchy labeling.

Finally, two fundamental challenges emerged: inherent AI distrust and limited solution-oriented content. One participant noted, \textit{``People using it for professional research are not necessarily very trusting of it, myself included'' (P4)} with ``it" referring to AI. Participants suggested linking original Reddit posts to build credibility. The scarcity of actionable solutions stems from the nature of informal forum discussions, which rarely propose concrete policy measures.

\section{DISCUSSION}
\subsection{\Sys Assists Policy Researchers Despite AI Skepticism}
Our evaluation of \Sys highlights its potential as a valuable complementary tool for policy research, offering unique opportunities to expand the scope of public opinion and streamline the research process. \Sys's report accurately covers the majority of themes present in the authoritative reports we studied, while supporting such findings with real-world anecdotes further demonstrating \Sys's potential as a supplementary tool. Interviewees also consistently emphasized that \Sys effectively captures diverse and unexpected insights from online forums -- perspectives that surveys and listening sessions often fail to include.

However, this satisfaction with \Sys’s outputs contrasts with a fundamental skepticism toward its AI foundation. This skepticism, noted in our evaluation, may reflect the demographic composition of our interviewees, which included several tenured professors and older policy researchers. If this skepticism is representative of the policy field as a whole, it could pose challenges to the broader adoption of \Sys and similar AI tools until their credibility is more widely established. However, the AI backend of \Sys exclusively generates results from real Reddit user posts, directly addressing concerns about transparency and authenticity.

\Sys exemplifies how AI can systematically integrate public discourse into policy research. By incorporating \Sys into research workflows, policy researchers can amplify diverse voices, especially those of internet-savvy individuals who may be absent from traditional listening sessions and surveys. Furthermore, its adoption could encourage greater engagement in online policy discussions, as individuals recognize their contributions may influence decision-making. Over time, integrating \Sys and similar tools into the research process could help build trust in AI, paving the way for broader acceptance of AI-assisted methods in policy research.
\subsection{Limitations and Future Work}
\label{sec:limitations}
\begin{itemize}[leftmargin=*,noitemsep]
    \item \textbf{Limited Study Duration.}
    In our user evaluations, the 10-15-minute time limit was set based on initial tests indicating theme saturation within that time and to respect interviewees’ schedules. We recognize that this time limit could be insufficient for a broader participant pool.

    \item \textbf{Improving Data Quality and Diversity.}
    One of the major concerns we received was the current focus on Reddit data that limited the scope of demographic representativeness and data quality. To account for this limitation, future work should have a greater breadth of pre-processed data sources provided in the backend, as well as allow data analysis on any uploaded dataset.

    \item \textbf{Credibility of Anonymous Online Forums.}
    Our current system's prompt focuses on extracting anecdotes from user data, rather than opinions to minimize bias. However, several participants noted that anecdotal evidence often contains misinformation, underscoring the credibility of the discussion. While it may not be feasible to fact-check all user data, filtering for accurately informed anecdotes would be a valuable next step.

    \item \textbf{Lack of Demographic Information.}
    The current absence of detailed metadata, such as user demographics or post geolocation, restricts a user’s ability to contextualize insights. Leveraging non-anonymous forums could allow more context behind user quotes to better inform policy researchers with deeper perspectives in their analysis. In addition to demographic context was a desire for quote context, such as quantity of engagement, upvotes, and comments.

    \item \textbf{Limited Quantitative Analyses.}
    Many users requested additional data visualizations and analyses leveraging the vast quantity of data collected over millions of discussions. Additional quantitative insights would greatly improve the value of the tool.

    \item \textbf{AI Distrust.}
    Many participants demonstrated inherent distrust in AI analysis and requested methods to bridge the trust gap. Suggestions included linking original posts to actual URLs to manually verify the legitimacy of quotes.
\end{itemize}
\section{Conclusion}
This work presents a novel approach to capturing public experiences related to policy-relevant issues by leveraging LLM-driven synthesis of online communities. Through our evaluation, we find that tools like \Sys can enhance policy researchers’ ability to identify diverse perspectives and rapidly gain thematic insights, ultimately complementing conventional methods such as in-person listening sessions and manual web searches. While \Sys currently draws from a limited data source—Reddit—its success points toward significant potential in scaling to broader online ecosystems. Expanding the platform’s corpus to include multiple social media channels and addressing limitations related to public data access would improve both the credibility and representativeness of its outputs.
In essence, \Sys represents an early step toward more inclusive, data-rich, and efficient policy analysis processes. By transforming sprawling online dialogues into actionable themes and validated quotes, our system has the potential to empower policy researchers and shape a more transparent, responsive, and democratic policy-making environment. As the platform matures and its reach broadens, we anticipate that tools like \Sys will become integral to how policymakers, researchers, and stakeholders engage with diverse, continuously evolving public conversations.

\begin{acks}
We thank Prof. Arvind Narayanan for helping define the project idea, Sayash Kapoor and Prof. Steven Kelts for their valuable feedback, and Kyler Zhou for helping with initial data analysis and prompt engineering.
\end{acks}
\bibliographystyle{ACM-Reference-Format}
\bibliography{references}

\clearpage
\appendix
\section{Appendix}
\subsection{Prompting \& Additional Technical Details}
\label{appendix:prompting}
Built on a Flask backend with LLM-powered analysis, the system processes social media content to surface both high-level themes and supporting evidence. To optimize performance, PolicyPulse caches processed reports for immediate future access and employs a modular architecture designed to scale across different data sources beyond its current focus on Reddit discussions. While Reddit isn't demographically representative of the general population, it is a public forum that cultivates voices from over 57 million daily active users and 138,000 active subreddit communities \cite{fiesler2024remember}. We embrace Reddit's core value to ``Remember the Human" by recognizing that our dataset represents real people's communications rather than mere data to be extracted. 

We used GPT-4 as our underlying LLM. The total cost to generate a complete report from 10,000 aggregated posts using PolicyPulse is approximately $\$150-$\$300, varying based on the number of quotes aggregated and the length of the posts. For prompting, our primary aim was to minimize human intervention during the text analysis pipeline, resulting in prompts structured to automatically generate high-level topic themes, categorize them into subtopics, and finally aggregate the results.

\subsubsection{Data Source Recommendation Prompt}
The current implementation of \Sys's data source recommendation system is specifically tailored to Reddit communities, serving as a proof-of-concept for our broader vision of public discourse analysis. The system uses the following prompt structure:

\begin{lstlisting}[basicstyle=\ttfamily\scriptsize]
Here is a list of subreddits: \{subreddits\_chunk\}. Based on the topic '\{topic\}', please provide a list of the most relevant 
subreddits from the list. If there are multiple relevant subreddits, separate their names with commas. If none are relevant, 
respond with a blank line.
\end{lstlisting}

While this implementation effectively serves our current focus on Reddit data, we designed the system's architecture with future extensibility in mind. The data source recommendation component is built as a modular system that can be expanded to incorporate diverse public data sources through the following architectural considerations

\subsubsection{Theme Generation Prompt}
\label{appendix:theme generation prompt}
For generating high-level themes, we developed a prompt that emphasizes policy relevance and structural consistency:

\begin{lstlisting}[basicstyle=\ttfamily\scriptsize]
Generate a list of 9 themes that policy researchers would be interested in learning more about, related to the subreddit 
\{subreddit\}, each with a title \{'title'} and a very brief description \{'description'}. Return the themes in JSON format.
\end{lstlisting}

The prompt’s design focuses on targeted audience alignment, specifically mentioned policy researchers, structural requirements, JSON output format for consistent parsing, and scope control setting a specific number of themes to ensure comprehensive coverage without overwhelming users.

\subsubsection{Processing Pipeline Stage 1: Quote Analysis Prompt}
The quote analysis prompt represents are most complex prompt engineering effort, designed to extract meaningful policy insights dynamically based on the users’ selected data source and provided theme:

\begin{lstlisting}[basicstyle=\ttfamily\scriptsize]
You are analyzing data from the subreddit \{\$subreddit}, containing discussions about various aspects of \{\$topic}, including 
post titles, introductory text, and main content. The high-level theme we're interested in is \{\$theme}. Your task is to extract 
only the most relevant quotes, personal experiences, and opinions that explicitly mention or discuss concerns, risks, or 
implications related to \{\$theme\_focus}.

Please process each row and output only quotes that:
1. Directly reference \{\$theme\_focus} concerns.
2. Address specific risks, dangers, or ethical concerns related to \{\$concerns\_scope}.
3. Include personal anecdotes or experiences discussing \{\$theme\_focus} implications of \{\$topic}.

For each relevant quote, create an output entry in JSON format with the following structure:
{
  "entries": [
    {
      "quote": "Full quote of a personal experience or opinion explicitly mentioning $theme_focus concerns in $topic",
      "summary": "A brief summary of the quote, providing context or the main idea in less than 8 words"
    },
    {
      "quote": "Another relevant quote or anecdote about $theme_focus",
      "summary": "Summary or context of this second quote in less than 8 words"
    }
  ]
}
\end{lstlisting}

\begin{quote}
If no quotes or relevant content about \texttt{\$theme\_focus} concerns are found in the data, return \texttt{null}.
\end{quote}

This is a templated prompt designed to adjust for varying subreddit forums and theme focuses. The context setting defines the data source and scope of the analysis, establishes the analytical framework, and specifies the theme focus to guide the overall direction of the project. Moreover, the filtering criteria require direct references to concerns related to the chosen theme, prioritizing specific risks, ethical considerations, and personal anecdotes or experiences that provide deeper insight into the topic. Lastly, the output structure includes the extraction of quotes with full context, concise summaries limited to under eight words, and the preservation of source references to maintain credibility and traceability.
\subsubsection{Processing Pipeline Stage 2: Subtopic Identification Across Aggregated Quotes}
For analyzing aggregated quotes and identifying subtopics from the aggregated data of relevant quotes and summaries aligned with the defined theme, \Sys employs a prompt focused on subtopic analysis for further fine-grained insights:
\begin{lstlisting}[basicstyle=\ttfamily\scriptsize]
You are a research assistant helping to analyze summaries of \{\$subreddit} discussions.

Analyze the provided summaries and identify the top 9 most prevalent themes or codes.  
For each code:
1. Provide a clear, concise name. The theme name should be specific and not too broad.
2. Provide a brief description.

Respond in valid JSON format with the following structure:
{
    "codes": [
        {
            "name": "Theme Name",
            "description": "Brief description of what this theme represents"
        }
    ]
}
\end{lstlisting}
\subsubsection{Processing Pipeline Stage 3: Quote Categorization Prompt}
The final stage involves mapping quotes to subtopics using a prompt that emphasizes precise categorization:
\begin{lstlisting}[basicstyle=\ttfamily\scriptsize]
You are a research assistant helping to categorize quotes about \{\$subreddit} discussions. You will be provided with:
1. A numbered list of codes (1-9) with their descriptions.
2. A list of quotes to categorize.

For each quote, assign the ONE MOST appropriate code number (1-9) based on the themes present in the quote.

Respond in valid JSON format with the following structure:
{
    "categorized_quotes": [
        {
            "quote": "original quote text",
            "source_id": "original source id",
            "codes": [
                {
                    "code": code_number,
                    "code_name": "name of the assigned code"
                }
            ]
        }
    ]
}
Guidelines for categorization:
- Assign THE ONE MOST relevant code to each quote.
- Include the theme that is most substantively discussed in the quote.
- Be consistent in how you apply the codes.
- Use the code descriptions to guide your decisions.
- The one code in the \texttt{codes} array should represent a significant theme in the quote, not just minor mentions.
\end{lstlisting}
Leveraging the 6 subtopics generated in the prior prompt, this prompt is designed to categorize quotes with subtopics consistently and accurately. It uses a number code system (1-6) with descriptions to standardize categorization and specifies a machine-readable JSON format for structured output. Particularly, the instructions emphasize selecting the one most relevant code based on substantive themes in the quotes, avoiding overcomplication or minor mentions. Clear guidelines ensure consistency, supported by the detailed code descriptions, enabling a streamlined and reliable process for thematic analysis.

\subsection{Detailed Screenshots of the \Sys Interface}
\label{appendix:screenshots}
\begin{figure*}[h]
    \centering
    \includegraphics[width=0.9\textwidth]{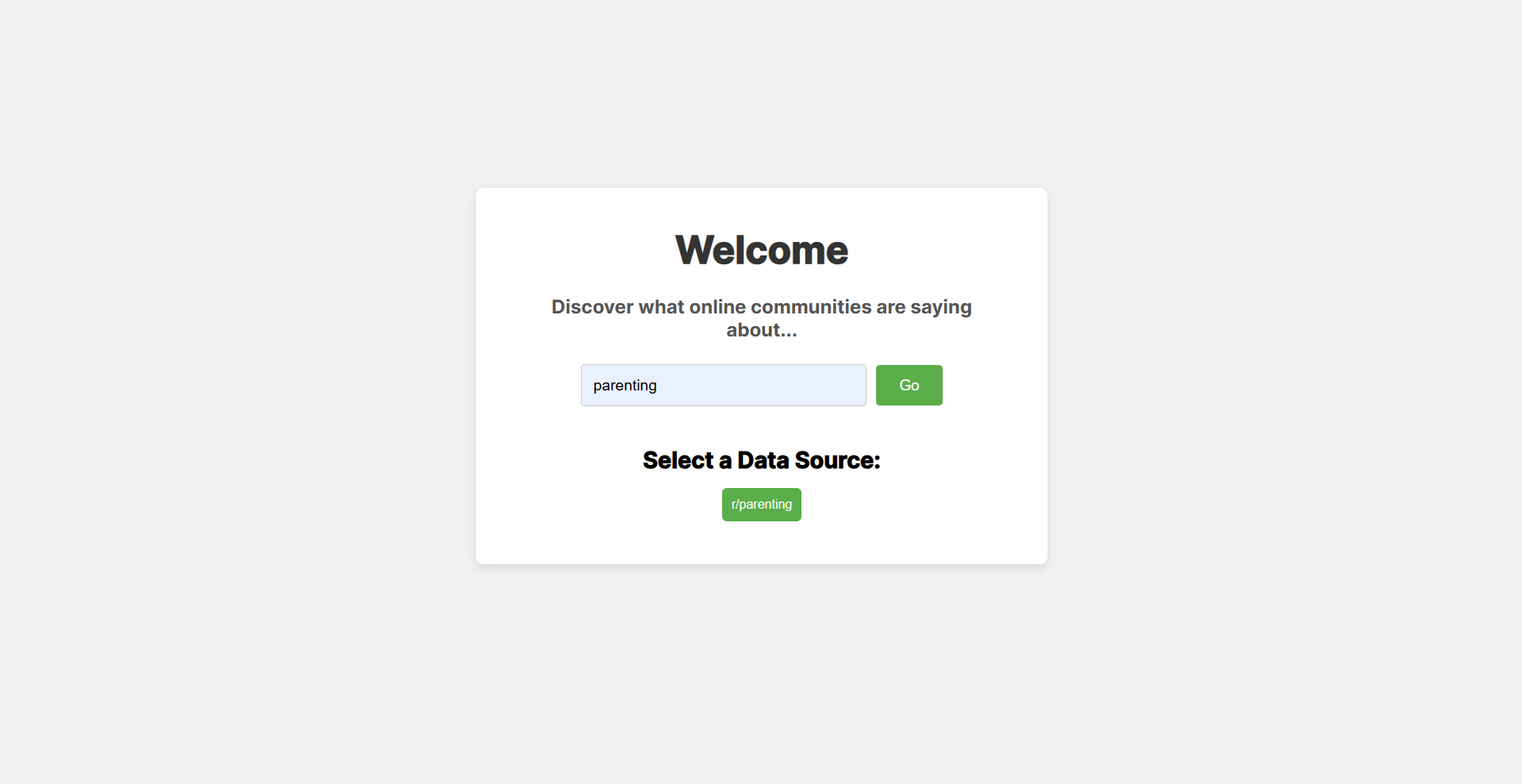} 
    \caption{Splash screen of the \Sys interface. This screen prompts users to input a policy for analysis. Based on the search query, \Sys will provide relevant data sources to choose from.}
    \label{fig:appendixSplash}
\end{figure*}

\begin{figure*}[h]
    \centering
    \includegraphics[width=0.95\textwidth]{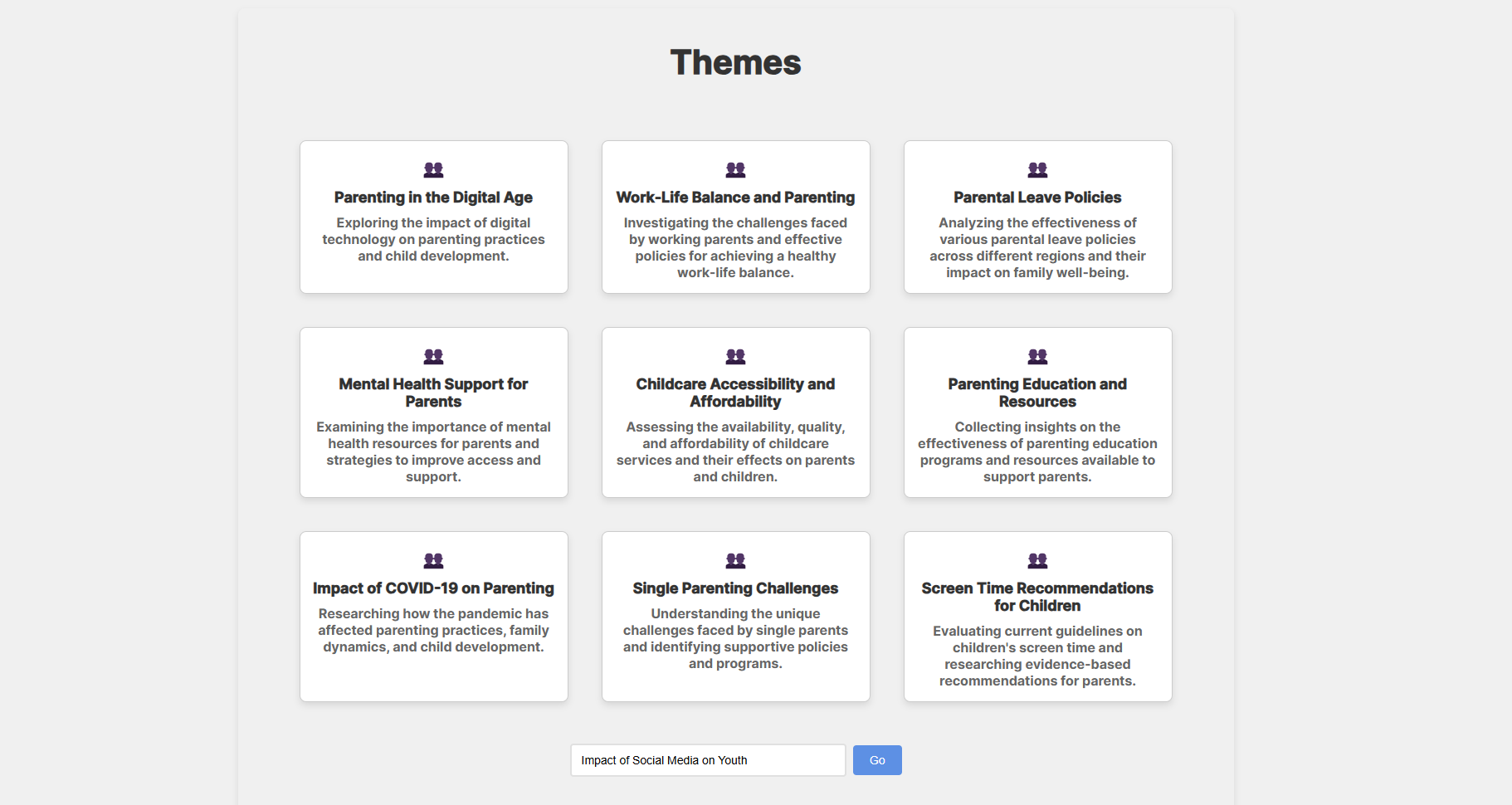} 
    \caption{Example of highlighted themes in the \Sys interface. After selecting a subreddit, users are presented with key themes from the discussion forum, each associated with real quotes.}
    \label{fig:appendixThemes}
\end{figure*}

\begin{figure*}[h]
    \centering
    \includegraphics[width=0.94\textwidth]{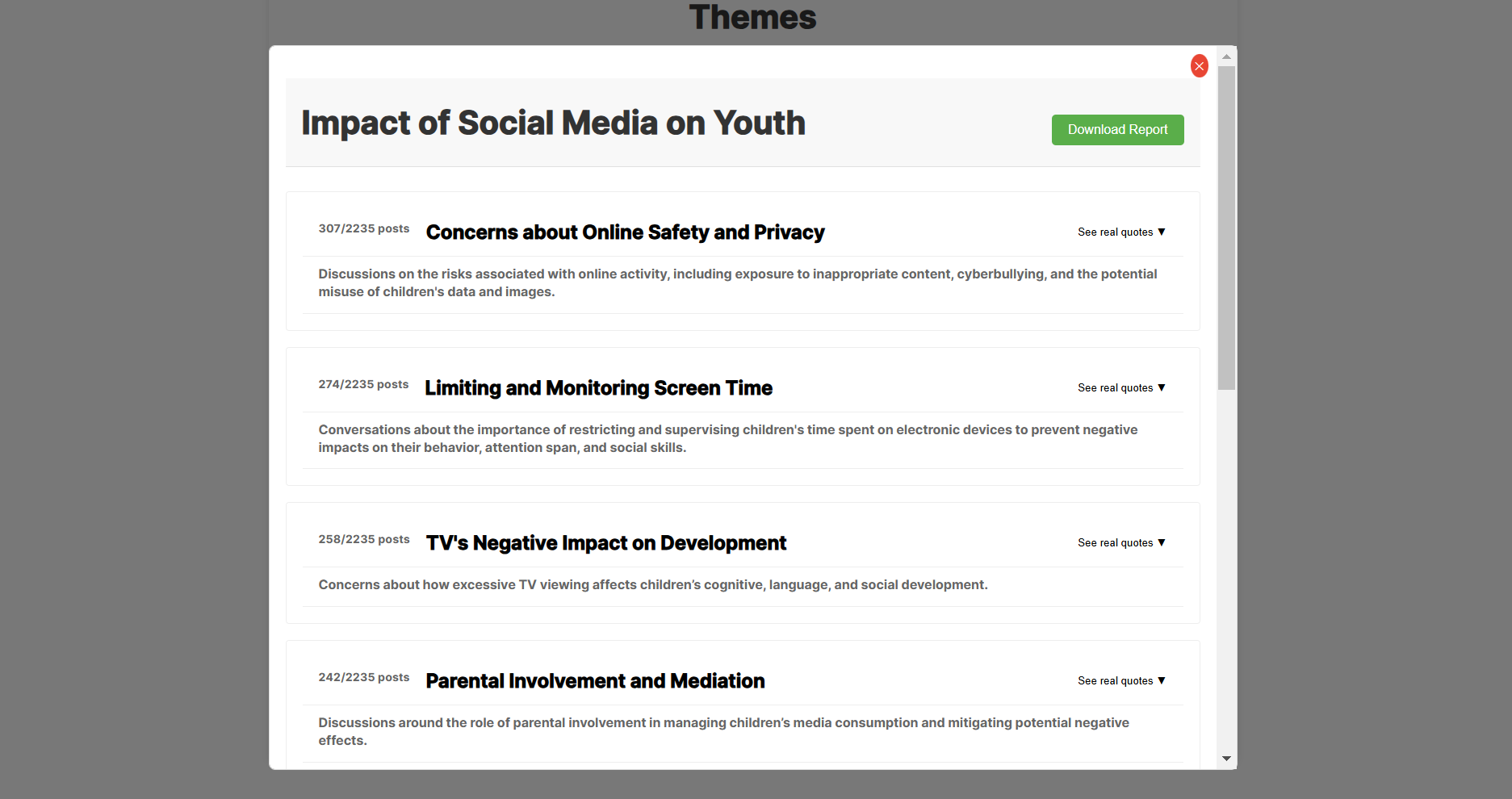} 
    \caption{Report screen of the \Sys interface. This screen displays key themes and opinions generated from subreddit discussions.}
    \label{fig:appendixReport1}
\end{figure*}

\begin{figure*}[h]
    \centering
    \includegraphics[width=0.94\textwidth]{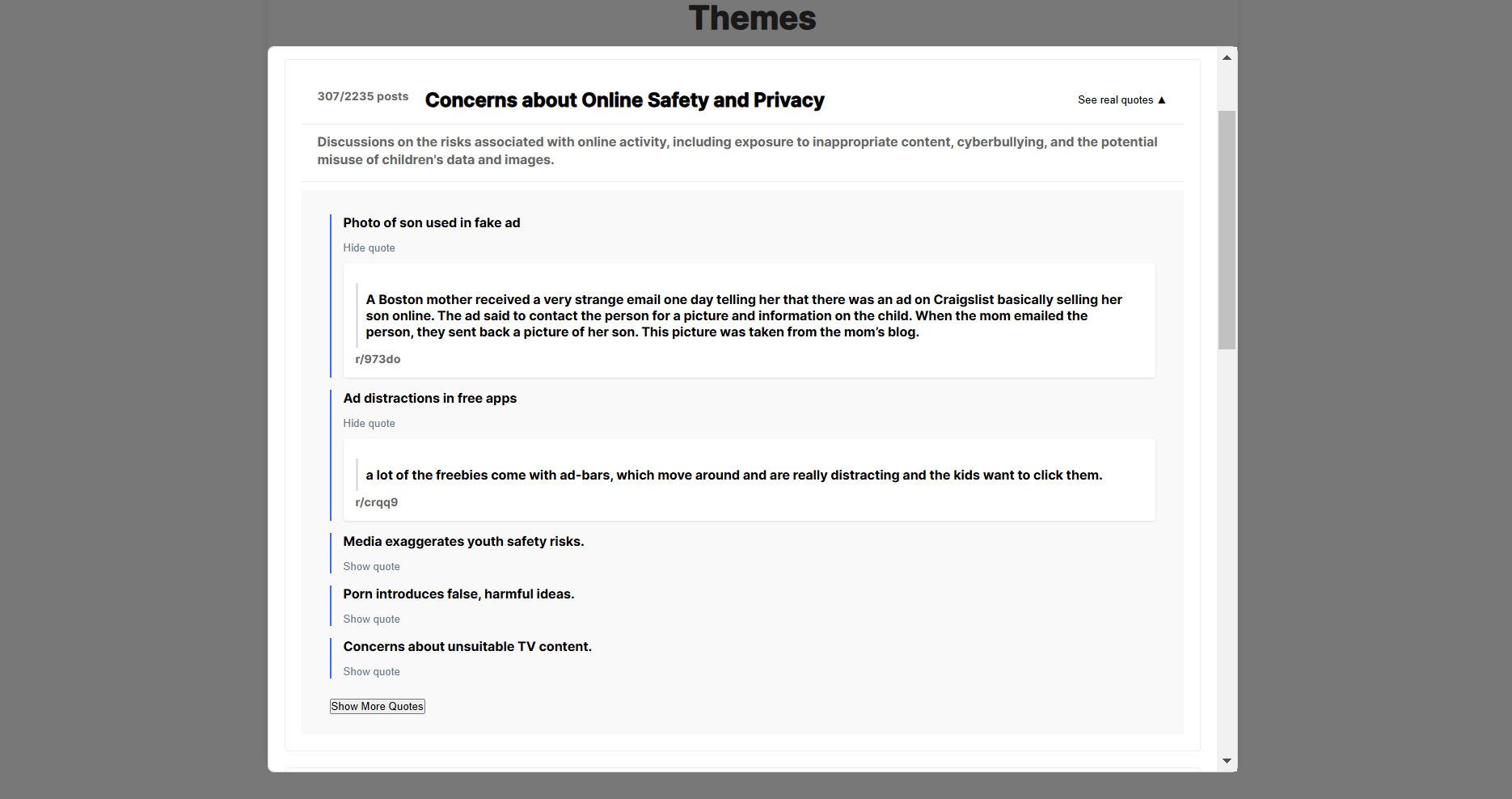} 
    \caption{Report screen of the \Sys interface. This screen displays key themes and opinions generated from subreddit discussions.}
    \label{fig:appendixReport2}
\end{figure*}

\clearpage
\subsection{Evaluation Script}
\label{appendix:evaluation}
\subsubsection{Pre-Task Questions} 
\begin{enumerate}
    \item How many years of experience do you have in policy research or analysis?
    \begin{itemize}
        \item In your current role, how often do you need to gather public opinion data? (Daily, Weekly, Monthly, Rarely, Never)
        \item How confident are you in your ability to quickly gather public opinions on a topic using online tools? (Scale 1-5, where 1 = Not confident at all, 5 = Extremely confident)
    \end{itemize}
    \item Topic Familiarity "Please rate your familiarity with the following topics:" (Scale 1-5, where 1 = Not familiar at all, 5 = Very familiar). How do you typically stay informed about these topics?
    \begin{itemize}
        \item Social Media Impact On Children
        \item Climate Change
    \end{itemize}
    \item AI Tool Experience
    \begin{itemize}
        \item Have you used any AI-assisted research tools before? 
        \begin{itemize}
            \item If yes, which ones?
            \item How frequently do you use AI-assisted research tools?
        \end{itemize}
        \item What are your expectations for how AI might help in gathering public opinion?
    \end{itemize}
    \item For your most recent policy research project:
    \begin{itemize}
        \item Roughly how many weeks did it take from start to finish?
        \item What percentage of that time was spent specifically on gathering public sentiment?
        \item Please sketch out your typical workflow when researching public opinion on a policy topic. Include all steps from initial research to final insights
        \begin{itemize}
            \item Estimate time spent on each step
            \item Mark which steps you find most time-consuming or complex
        \end{itemize}
    \end{itemize}
\end{enumerate}

\subsubsection{Task Procedure}
Phase 1 (10 minutes): "You'll be assigned to either Topic A (Social Media and children) or Topic B (Climate Change). Your goal is to gather as many meaningful insights about public opinion on this topic as possible in 10 minutes. Try to identify both specific anecdotes and broader themes in public discourse.

[For AI-assisted group]:
\begin{itemize}
    \item You'll be using an AI research tool, which can help analyze social media posts and online discussions
    \item The tool will show you relevant posts and help identify common themes
    \item You can use the tool's features to filter and categorize the information
    \item Please take notes on any insights you find using the given template
\end{itemize}

[For traditional method group]:
\begin{itemize}
    \item You'll have access to standard web browsers and search tools
    \item You can visit any public websites, forums, or social media platforms
    \item You may not use any AI tools such as ChatGPT
    \item Please take notes on any insights you find using the given template
\end{itemize}
Remember to focus on both individual stories and broader patterns in public opinion."
Phase 2 (10 minutes): "Now you'll switch to the other topic and use the other research method. The same goals apply - gather as many meaningful insights as possible in 10 minutes."

\subsubsection{Post-Task Survey}
"Now that you've completed both tasks, please share your experience with each method."
Quantitative Metrics:
\begin{enumerate}
    \item For each topic and method, please indicate in the given template:
    \begin{itemize}
        \item Number of distinct anecdotes gathered
        \item Number of broader themes identified
        \item Time spent finding your first meaningful insight
        \item Confidence in the comprehensiveness of your findings (Scale 1-5
    \end{itemize}
\end{enumerate}

\textbf{Process Comparison}
“Now that you have used the AI tool, sketch your research workflow using \Sys for the same research task." 
\begin{itemize}
    \item Write the new process  (real quotes and testimonies) 
    \item Mark steps that were:
    \begin{itemize}
        \item Eliminated completely
        \item Simplified/compressed
        \item Unchanged
    \end{itemize}
    \item For simplified steps: Briefly explain how \Sys changed the step
\end{itemize}

\textbf{Comparative Questions (5-point scale)}
Rate how much faster or slower \Sys was for:
\begin{itemize}
    \item Finding relevant public opinions (Much slower 1-2-3-4-5 Much faster)
    \item Understanding sentiment trends (Much slower 1-2-3-4-5 Much faster)
    \item Generating summary reports (Much slower 1-2-3-4-5 Much faster)
\end{itemize}
Which method was more effective for:
\begin{itemize}
    \item Finding specific examples/anecdotes?
    \item Identifying broader themes?
    \item Understanding the range of public opinions?
    \item Discovering unexpected insights? (Rate each on a scale 1-5 for both methods
\end{itemize}

User Experience: For the AI-assisted tool:
\begin{itemize}
    \item How intuitive was the interface? (Scale 1-5)
    \item What features were most helpful?
    \item What features were missing or could be improved?
    \item Did you trust the tool's analysis? Why or why not?
\end{itemize}

For the traditional method:
\begin{itemize}
    \item What strategies did you use to find information quickly?
    \item What were the main challenges you encountered?
    \item What advantages did this method have over the AI tool?
\end{itemize}

Comparative Analysis: Please compare the two methods:
\begin{itemize}
    \item Which method would you prefer for future research tasks? Why?
    \item How did the effectiveness of each method vary between the two topics?
    \item What would be your ideal combination of AI-assisted and traditional research methods?
\end{itemize}

\subsection{Interview Worksheet} \label{sec:Interview Worksheet}
Topic: [Social Media’s Impact on Youth/Climate Change]. Method: [Traditional or AI-Assisted]

\textbf{Themes Identified}
\begin{enumerate}
    \item Theme: 
    \begin{itemize}
        \item Brief Description:
        \begin{itemize}
            \item Describe the main stakeholder perspectives
            \item Summarize proposed solutions or interventions mentioned in the discussions
            \item Note any emerging trends
            \item Note significant disagreement or consensus
        \end{itemize}
        \item Supporting Evidence (2-3 examples): 
        \begin{itemize}
            \item Can be an anecdote/story
        \end{itemize}
    \end{itemize}
        \item Theme: 
    \begin{itemize}
        \item Brief Description:
        \item Supporting Evidence (2-3 examples): 
    \end{itemize}
\end{enumerate}
[Continue for each theme...]

\textbf{Quick Statistics}
\begin{itemize}
    \item Total Themes Identified: [Count]
    \item Total Anecdotes Collected: [Count]
    \item Time to First Insight: [Minutes]
\end{itemize}

\textbf{Notes}
[Any quick observations about the research process or findings]

\onecolumn
\subsection{Comparison to Authoritative Results Details}
\label{appendix:auth_comparison}

\subsubsection{Climate Change}

 PolicyPulse Themes:
 \begin{enumerate}
 \item The Effects of Climate Change on Water Systems
\item Climate-Related Health Impacts
\item Economic Factors Driving Climate Change
\item Environmental Justice and Equity
\item Global Biodiversity Loss
\item Causes and Effects of Global Deforestation 
\item Indigenous Communities and Climate Vulnerability
\item International Climate Policies and Agreements
\item Land Degradation and Desertification
\end{enumerate}

\begin{table*}[h]
    \centering
    \small
    \begin{tabular}{p{7cm} | c| c} 
        \toprule
        \textbf{Themes in Authoritative Report}& \textbf{Present in PolicyPulse (Y/N)?}& \textbf{PolicyPulse Theme (Numeric)}\\
        \midrule        
        Atmospheric Composition and Global Climate Drivers& Y & 6,9\\
        Ocean System Transformations& N & X\\
        Cryosphere and Polar Region Changes& N& X\\
        Extreme Weather and Climate Events& Y& 6,9\\
        Precipitation and Water Cycle Dynamics& Y & 1\\
        Ecosystem and Biodiversity Responses& Y& 5\\
        Climate and Human Health Impacts& Y& 2\\
        Food Systems and Agricultural Resilience& Y& 9\\
        Human Mobility and Social Vulnerability& Y& 7\\
        Climate Policy and Sustainable Development& Y& 8\\
        Economic and Infrastructure Adaptation& Y& 3\\
        Climate Change Technology and Innovation& N& X\\
        Natural Resource Management& Y & 4\\
        Climate Communication and Public Awareness& Y & 4\\
        Interdisciplinary Climate Research& N& X\\   
        \bottomrule
    \end{tabular}
    \caption{Theme Comparison across the World Meteorological Organization’s authoritative report, “The Global Climate 2011–2020: A Decade of Accelerating Climate Change" and \Sys.}
    \label{tab:policy_comparison1}
\end{table*}


\subsubsection{Digital Parenting in the Age of Technology}
\Sys’s report (before expansion to include quotes) on “Digital Parenting in the Age of Technology” contains the following themes:
\begin{enumerate}
    \item Internet safety for children
    \item Concerns about excessive screen time
    \item Parental control and monitoring
    \item Exposure to inappropriate content
    \item Balancing technology and other activities
    \item Digital responsibility and accountability
    \item Tech’s impact on mental health
    \item Social media and children’s privacy
    \item Gaming addiction concerns
\end{enumerate}

Shown in Table~\ref{tab:policy_comparison2} are the themes present in the Pew Research Center's authoritative report on “Parenting Children in the Age of Screens” and whether they are present in \Sys’s report. The themes were collected from the authoritative report through thematic coding.

\begin{table*}[htb]
    \centering
    \small
    \begin{tabular}{p{8cm} | c | c}
        \toprule
        \textbf{Themes in Authoritative Report} & \textbf{Present in PolicyPulse (Y/N)} & \textbf{PolicyPulse Theme (Numeric)} \\
        \midrule
        Increased Concern About Screen Time & Y & 2 \\
        Parental Anxiety Over Technology Use & Y & 4 \\
        Smartphone and Tablet Usage Among Young Children & Y & 3 \\
        Parental Regulation of Screen Time & Y & 2 \\
        Impact of Smartphones on Social Skills and Friendships & Y & 9 \\
        Use of YouTube Among Young Children & N & X \\
        Exposure to Inappropriate Content & Y & 4 \\
        Parental Digital Distraction & N & X \\
        Diverging Views on When Children Should Have Their Own Devices & Y & 3 \\
        Concerns About Online Safety & Y & 1 \\
        The Role of Schools and Doctors in Digital Guidance & Y & 7 \\
        Parental Monitoring of Digital Activity & Y & 3 \\
        Technology’s Impact on Education & Y & 5 \\
        The Growing Prevalence of Voice Assistants & N & X \\
        Parental Enforcement of Digital Grounding & Y & 9 \\
        Racial and Socioeconomic Differences in Screen Time Views & Y & 2 \\
        The Role of Social Media in Children's Lives & Y & 8 \\
        Technology as a Parenting Challenge & Y & 3 \\
        Debate Over Who Should Regulate Children’s Digital Experiences & Y & 6 \\
        \bottomrule
    \end{tabular}
    \caption{Theme Comparison across the Pew Research Center’s authoritative report, "Parenting Children in the Age of Screens" and \Sys.}
    \label{tab:policy_comparison2}
\end{table*}

\end{document}